# WIRELESS LAN TO SUPPORT MULTIMEDIA COMMUNICATION USING SPREAD SPECTRUM TECHNOLOGY.

**Sourav Dhar, Rabindranath Bera, K.Mal**

***Sikkim Manipal Institute of Technology, Sikkim Manipal University, Majitar, Rangpo, East Sikkim 737132, INDIA, E-mail: sourav80@indiatimes.com; r_bera@hotmail.com***

## ABSTRACT

Wireless LAN is currently enjoying rapid deployment in University departments, business offices, hospitals and homes. It becomes an inexpensive technology and allows multiple numbers of the households to simultaneously access the internet while roaming about the house.

In the present work , the design and development of a wireless LAN is highlighted which utilizes direct sequence spread spectrum (DSSS) technology at 900MHz RF carrier frequency in its physical layer. This provides enormous security in the physical layer and hence it is very difficult to hack or jam the network. The installation cost is also less due to the use of 900 MHz RF carrier frequency..

## INTRODUCTION

This project is proposed to implement mobile virtual private network (MVPN) in an intelligent transport system (ITS). As a first step to this, the laboratory work is done to establish a wireless LAN system which will be able to support multimedia communication. i.e, the network should support the transmission and reception of voice, data and image. In virtual private network, security is most important. To provide enough security spread spectrum technology has been used in the physical layer. For software development MATLAB (version 6.5) has been used.

In this work the data and voice transmission and reception have been successfully tested. Here the data has been assigned higher priority than voice, i.e, whenever data and voice transmission will be attempted simultaneously, the data will be transmitted suppressing the voice. This leads to an “either or “situation. Transmission and reception of both data and voice have been done using the audio port of the computer. For the transmission and reception of data “chat window “is established using software.

We are using spread spectrum(SS) technology because the SS technology is well known for its anti jamming and security features and has the following number of advantages[1], [2].

a) Frequency diversity : Because the carrier signal is spread out over a larger bandwidth, frequency-dependent transmission impairments , such as noise bursts and selective fading , have less effect on signal.

b) Multi path resistance: In addition to the ability of SS to overcome the multi path fading by frequency diversity , the chipping codes used for CDMA not only exhibit low cross correlation but also high auto correlation. Therefore, a version of the signal that is delayed by more than one chip interval does not interfere with the dominant signal as much as in other multi path environments.

c)Security : Because spread spectrum is obtained by the use of noise like signals , where each user has a unique code , security is inherent .

d)Anti jamming: The noise like spread spectrum helps a lot in providing resistance to a anti jamming radio source.

To establish a communication system properly, two steps has to be followed. First the connection is to be established between the transmitter and receiver (handshaking), then original information can be conveyed (communication). Until and unless the connection is established we can’t transfer the information. Thus there are two parts of this project. First is handshaking and second is communication[3], [4].

## ARRANGEMENT FOR WIRELESS COMMUNICATION BETWEEN TWO COMPUTERS

In order to establish connection between two of the desired computers, the transmitter as well as the receiver both should have a codec. Fig.1 shows the arrangement for wireless communication system to be connected between two computers . These wireless units consist of a codec, which is connected to the DSSS radio unit , a computer and a zero-crossing detector. After handshaking, communication is done using the humming bird ASIC of DSSS set which is connected to audio port of the computer.

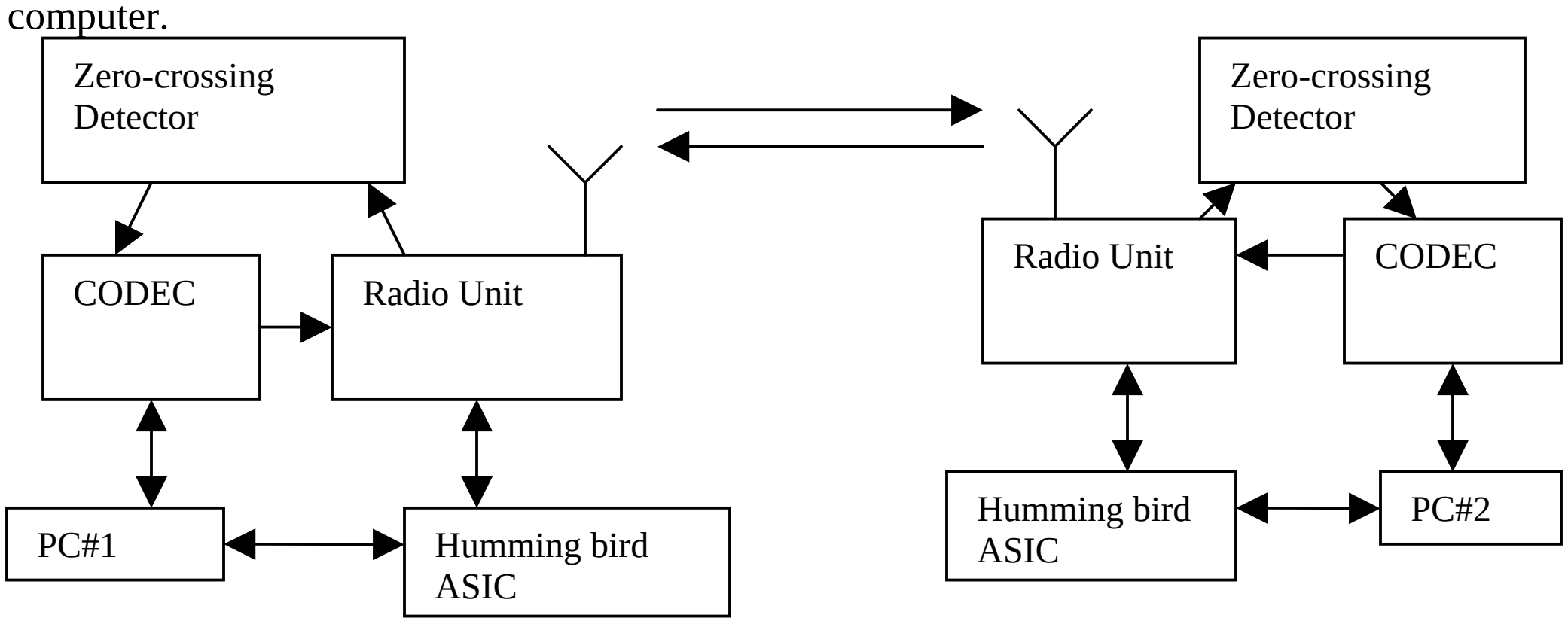

Fig.1: Arrangement for wireless communication between two computers.

## HANDSHAKING

As shown in the fig.2 there is a switch which enables the CODEC which in turn generates the desired code and also the transmit enable signal. The receive enable pin of the radio unit is always high. The transmit enable signal from the codec enables the transmission of the radio unit. Then the code is transmitted. The code frame contains a start bit, followed by 6 bit source address, then 6 bit destination address and finally an acknowledgement bit. Initially the acknowledgement bit is made ‘0’.Requested code from a mobile unit is broadcasted by the base station and will be received by all users. All users in turn will decode the requested code and after decoding, they will co-relate the code. Now only that user, whose correlation coefficient is greater than 90% will retransmit the same code by exchanging source and destination address with acknowledgement bit high.  For other users the correlation coefficient is in the order of 0% and hence they will discard that code. Base station will forward the code to the sender through broadcasting and the connection will be established in this fashion between source and destination. Now “ TALK” mode will be written on the LCD display of the sender. The multi media communication will begin next which is PC based, i.e; now data, voice or video can be transferred.

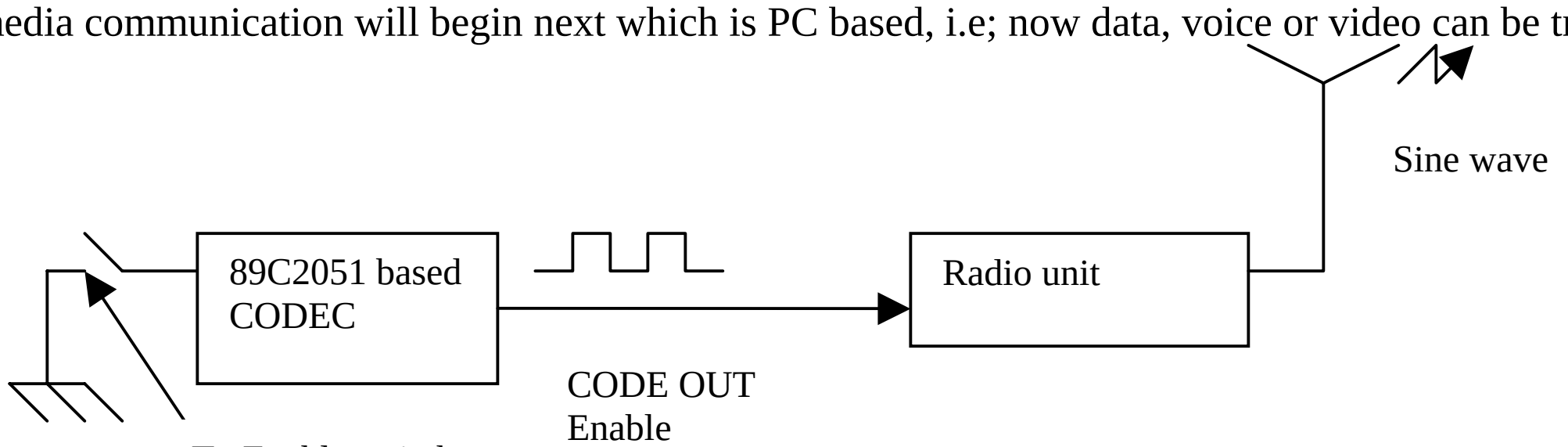

Fig.2: experimental setup for handshaking

## Concept of codec program

For handshaking, the source will transmit a code consisting source & destination address, a start bit and an acknowledgement bit. This is a 14 bit code and generated by software program. The code format is given in fig.3.

| | | | |
|---|---|---|---|
| Start bit | 6bit destination address | 6bit source address | ACK bit |

14 bit

Fig.3: code format (14 bit)

First start bit is made 1 for enabling transmission. And the acknowledgement bit is 0 for the transmitter wish to communicate. On reception of this code the destination will interchange the source and destination addresses and will make acknowledgement bit 1, and then retransmit the code. Hence the connection will be established between two computers.
As shown is fig.4, 1 is coded by a pulse having pulse repetition time (PRT) 800µs and zero is coded by a pulse having PRT 600µs. these pulses are of 50% duty cycle. And pulse amplitude is made 20mv.
Now let us take an example to make it clear. Computer 8 wishes to communicate with computer 1. Then the source address will be 001000 and destination address will be 000001. The source address is stored in a register of the microcontroller. There is a lookup table where the addresses of the other computers are stored. Thus on reception of a start command the codec (microcontroller) of the transmitting computer8 generates a 14 bit code with acknowledgement bit 0.

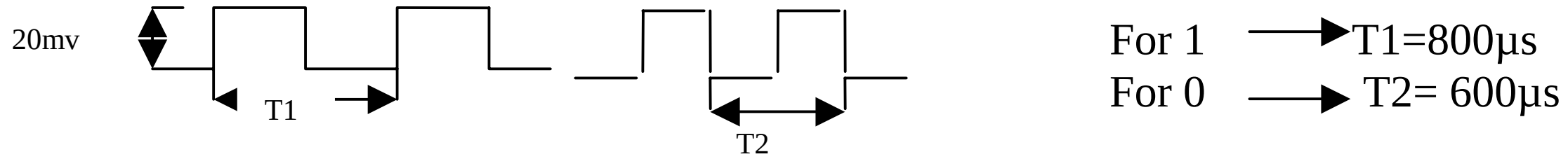

Fig.4: pulses for corresponding codes 0 & 1.

On reception of this code, codec of computer 1 will first check the destination address , if it matches with its address that is already stored in its one of the registers, then it interchanges the source and destination addresses and makes acknowledgement bit 1 and then retransmits the code. This code is transmitted serially i.e. in a sequential fashion.
As this radio unit uses frequency shift keying (FSK) modulation, thus this code will be FSK modulated in the radio unit at 900MHz and then transmitted. At the output of the radio unit we will have compulsorily a sine wave. Same sinusoidal analog signal will be received by the radio unit of the computer 8. But the codec cannot deal with the sinusoidal signals. Thus to decode the received signal, it should be converted into square wave first to provide to discrete level. The simplest of the sine to square wave converters is a comparator used as a zero-crossing detector. The output of zero crossing detector is shown in fig.5.

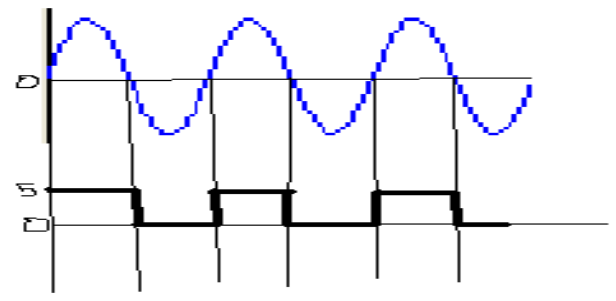

Fig. 5: out-put of zero-crossing detector (pulse recovery)

Thus pulse recovery is done. But pulse recovery is not code recovery. The code recovery is done by actual measurement of pulse repetition time (PRT). And a software based decision will be taken by the micro controller. Block diagrammatic representation of code recovery is shown in fig.6

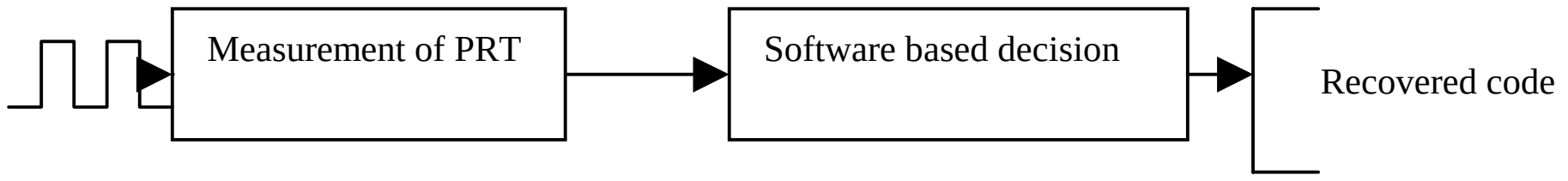

Fig.6: code recovery

This code recovery is done within the microcontroller. Hence this is entirely software based. There is a counter in the microcontroller which is enabled by the negative edge trigger. As soon as a negative edge comes in , the counter will start counting it will count till another negative edge comes in. the count value will be latched and stored into one of the registers. Time for one count will be decided by the frequency of the crystal that is used for providing clock to the microcontroller. After measurement of PRT , the result is given to the decision device. A threshold level is set there, in this case we set threshold at 700µs. if the PRT is less than 700µs then it is zero and if it is more than 700µs then it is 1. thus the code is recovered.

**COMMUNICATION**

In this project we are trying to first establish multimedia communication between two computers. To achieve this we have purchased a spread spectrum cordless telephone set. The base unit of this telephone set is connected to one computer and the hand set is also connected to another computer. The interfacing between the computer and the telephone set is done using the audio port and the microphone input of the computer. Both of the base unit and the hand set consists of humming bird application specific integrated circuits (ASIC). This humming bird device provides higher integration, improved audio quality and lower power consumption. Hummingbird consists of an ASIC into which a base band modem an audio modem and controller are integrated and a linear audio CODEC. The base band modem provides all kind of modulation and demodulation, all sort of encoding and decoding and the Spreading operation. The audio modem consists of a 40 Kbps ADPCM engine that interfaces to the audio CODEC and a built in DTMF detector. The controller is an embedded MC19 microcontroller. Job of this is monitoring the different functions of the DSSS telephone. The audio CODEC converts the analog signals from the PSTN and microphone to and from digital voice samples for the audio modem. It consists of built in electronic microphone interfaces and Independent audio channels for line and speaker interfaces. Along with the hummingbird ASIC the base unit and the hand set also radio unit (RF 105). This radio unit is taking care of the transmission and reception at 900MHz. RF 105 , a fully integrated transceiver device. It provide transmit, receive and frequency synthesis function for digital spread spectrum (DSSS) system operating in 902 to 928 MHz industrial, scientific and medical (ISM) band.

For software development we are using MATLAB (version 6.5). Using this software we have established the “chat mode”. So the arrangement supports more user friendly approach. So far only alphanumeric characters have been coded. Coding is done using dual tone multiple frequency (DTMF) encoding scheme. This approach supports us to use the audio ports for interfacing the telephone set with computers.

In the software part, a look up table is generated where the two frequencies corresponding to each character are listed with a tolerance of ±5%. The same program is used in transmitter side as well as in the receiver side. The chat window is established using graphic user interface (GUI) of MATLAB-6.5.

At the transmitter side, the character to be sent is typed in the transmit window. Comparing the character with the characters of the lookup table, corresponding frequencies are generated using software and then transmitted through the audio port of the computer.

At the receiver side, after receiving the signal through the microphone input of the computers, its FFT is done. Peaks of the spectrum are detected and the frequencies of the peaks of the spectrum are estimated online. Then these frequencies are compared with the frequencies of the look up table and the corresponding characters are displayed in the receive window that shown above.

## CONCLUSION

It is a well established fact that spread spectrum based wireless technology is coming with great hope to support user need in the form of multimedia service compared to existing technology. It has started to support high quality voice in mobile environment without any multimedia service. We are trying to exploit this fact towards the multimedia communication. Due to its high anti jamming quality spread spectrum technology is essential in the mobile computing field for its reliable operation.

## ACKNOWLEDGEMENT

The authors are thankful to the technical assistants of Electronics & Communication Engineering Department, SMIT, and Prof. K.K.Gupta, Dean, SMIT for providing them support both technically and financially.

## REFERENCE

[1] “Code Division Multiple Access -A Modern Trends in Multimedia Mobile Communication”.R. Bera,S. K. Sarkar, D.Kandar, J. Bera , S.S. Singh, S. Dogra , *ICSM, 2005 held at IIMT, Gurgaon, Delhi during March 11-12.*

[2] “Digital communication systems”, Simon Haykin, *John Wiley & Sons,[2004]*

[3] “ Computer networking –a top down approach featuring the internet”, J.A.Kurose and K.W.Ross, *Pearson Education Asia publication, pp 480-487,2003*

[4] “Wireless communication and networks”, W. Stallings, *Pearson Education Asia publication, pp 320-333,[2002]*